\documentclass{article}
\usepackage{graphics}
\usepackage[dvips]{graphicx}
\usepackage{bm}
\usepackage{latexsym}
\usepackage{pxfonts}
\begin{document}
\title{Signature of multiple glassy states in micellar nanoparticle-polymer composites}
\author{A.K.Kandar$^{1}$, S. Srivastava$^{1}$, J. K. Basu$^{1\ast}$, M.K. Mukhopadhyay$^{2}$, S. Seifert$^3$ and S.Narayanan$^{3}$\\\\\ 
\normalsize{$^{1}$Department of Physics, Indian Institute of Science ,Bangalore, 560 012 India,}\\
\normalsize{$^{2}$Department of Physics, University of California San Diego, La Jolla,  CA 92093, USA}\\
\normalsize{$^{3}$Advanced Photon Source, Argonne National Laboratory, Argonne, IL 60439, USA.}}
\maketitle
\begin{abstract}
We present results of temperature dependent measurements of  dynamics of micellar nanoparticle - polymer composites of fixed volume fraction and variable polymer chain grafting density. For nanoparticles with lower grafting density we observe dynamically arrested state at low temperatures corresponding to an attractive glass while at high temperature the same system shows relaxation typical of a repulsive glass. For higher grafting density, the low temperature dynamics resembles more of a gel which crosses over to a repulsive glass at high temperature. Possible reasons for such fascinating dynamical transitions is delineated.
\end{abstract}
The physics of soft materials is quite intriguing due to the existence of interplay of various competing interaction at different length and time scales [1-2]. This has led to extensive investigation  of the thermal, mechanical and rheological properties of such systems through experiments,simulations and theoretical calculations [3 - 6]. One of the prototypical system belonging to this class of systems is colloidal particles which are known to undergo both gelation as well as glass transition under suitable set of conditions [2,7 - 11]. For a long time these two phenomena treated as independent both in experiments as well as theory. However recently an attempt has been made to unify these two apparently dissimilar concepts into the frame work of a single theory-namely the Mode coupling Theory(MCT) of liquids and glasses [12-14]. It has been shown that by tuning the interaction potential  between colloidal particles one is able to observe multiple glassy states and re-entrant glass liquid transitions [7-9,15-18]. It turns out that for this to happen the particles must interact through not only a strong hard core repulsion but also a short-range attraction. By tuning the range and depth of the potential of attractive interaction it is possible to drive a colloidal system from a repulsive to an attractive glassy state.It is predicted that the transition to a gel state from a liquid is seen to occur at a low volume fraction while the transition to an attractive or repulsive glass is found to occur at considerably higher volume fractions [16]. Recently this has also been verified experimentally [19]. A crucial prediction of the MCT for such systems is the existence of a region of logarithmic relaxation. Experimentally, such predictions have been verified for colloidal systems [16 - 17] as well as for polymeric micellar systems [15] by tuning the strength of the interaction through addition of polymers or changing temperature.  Nanoparticle - polymer hybrid materials also share some of the rich and complex thermo - mechanical behavior of soft, glassy systems [3,20-21]. For instance, it has been shown that addition of nanoparticles into a polymer matrix can lead to reduction of viscosity with increase in volume fraction instead of the expected increase according to Einstein relation [22]. The most effective way of dispersion of nanoparticles in polymer matrices is to control the particle - polymer interface and a very efficient method to ensure this is to graft nanoparticles with polymer which is
chemically analogous to the matrix polymer [21,23-26]. It is now clear that the polymer - particle interface is the most crucial parameter in determining the various thermal, mechanical and related properties of the hybrid system[3,20-21,25-26]. Interestingly, the nanoparticles with  high polymer grafting density leads to the formation of a star like polymer particle or a dense polymer brush with a hard core. The thermal properties of such composite particles, in the absence of solvent, and especially the temperature dependence of their dynamics on the grafted chain conformation, should provide valuable insight into the observed thermo - mechanical and anomalous rheological behavior of nanoparticle - polymer hybrid systems [3,20-22,25-26].In this letter ,we study the dynamics of (solvent free) polymer grafted composite nanoparticles using temperature dependent x- ray photon correlation spectroscopy (XPCS). The dynamics of such composite particles revealed strong similarities with other related soft materials in the form of displaying the existence of both a repulsive as well as an attractive glassy phase. The volume fraction ($\phi$) of such composite particles is $\sim$ 1 (solvent free) and at such high fractions the differences between the two glass phases should ceased to exist as per the predictions of MCT. We show for the first time how the glass - to - glass transitions is mediated by morphology of grafted polymer chain which can be tuned by controlling the grafting density and temperature. \\
The Micellar nanoparticles were synthesized using a modified version of the method used in [27]. Gold nano particles of controlled sizes and interface morphology were synthesized by the well known method of
reduction of gold chloride (0.8mM) in tetrahydrofuran (THF)  with
superhydride (0.1M) in the presence of thiol-terminated polystyrene
(PST) (0.16mM) of molecular weight 53K (Polymer Source). The excess
PST was removed by precipitation of the PST capped gold nanoparticles from solution by adding ethanol. The last process was repeated several times to ensure that almost no free PST chains were present. Then PST capped nanoparticles were redissolved in chloroform for Ultra Violet-Visible (UV - Vis) absorption spectroscopy of the PST capped gold nanoparticles confirmed formation of very small nanoparticles and the sizes of the nanoparticles were further confirmed using transmission electron microscopy (TEM) revealing formation of fairly monodisperse nanoparticle cores. The final powders of these nanoparticles were extracted from solutions by precipitation followed by heating at 50$^0C$. It is expected that the strong gold-thiol bonding should lead to formation of stable gold nano particles with the PST chains grafted on their surface through the thiol-end of the respective chains. Incoherent small angle x - ray scattering (SAXS)measurements on powder of both the samples were also performed at beamline 12 - ID of APS with an incident beam energy of
12KeV. XPCS experiments were performed at the Advanced Photon source (APS) in Argonne National Laboratory (ANL) at the 8 - ID beam line. All measurements were performed with a monochromatic beam of 7.35 KeV  x - rays. The incident beam on the sample was collimated to 20 $\mu$ x 20 $\mu$m to obtain a partially coherent beam of radiation and the scattered beam was recorded on  CCD camera (Princeton Instruments). The powder samples were annealed at 423K for 24-30 hours under a vacuum of 5 $\times$ 10$^{-4}$  mbar or better before XPCS measurements. In XPCS one measures the intensity autocorrelation function [18,28],
{\begin{equation}
g_2(q,t)= 1 + b{\vert F(q,t)\vert}^2.
\end{equation}}\
Here, $F(q,t)$ is the intermediate scattering function (ISF) and $ b $
is the speckle contrast and $q$ is the measured wave vector. XPCS measurements are reported for two such samples A and B to bring out the dramatic difference in dynamics with respect to various parameters like,the glass transition temperature T$_g$ or grafting density, $ \sigma $. The size and morphology of the nanoparticles could be controlled by varying gold-polymer relative fraction as well as reaction time as described ealier[27]. In this case our goal was to obtain nanoparticle with variable capping density, $ \sigma $ keeping the size as similar as possible for which we varied the reaction time only. Sample A consists of gold nanoparticles of core diameter 2.4 nm with a dense packing of PST. The density of capping chains was estimated to be $\sim$ 0.4 chains/nm$^{2}$ from thermo gravimetric analysis (TGA) measurements and the core volume fraction is $\sim$ 2$\%$. The T$_g$ of this sample was estimated to be 380K from modulated differential scanning calorimetry experiments. Sample B consists of 3.2 nm core diameter gold nanoparticles but with a higher PST capping density of $\sim$ 2.3 chains/nm$^{2}$ which was achieved by carrying out the reduction reaction for larger time. The core volume fraction is $\sim$ 0.4$\%$. The T$_g$ of this sample was found to be 370K. 
\begin{figure}
\includegraphics[scale=2]{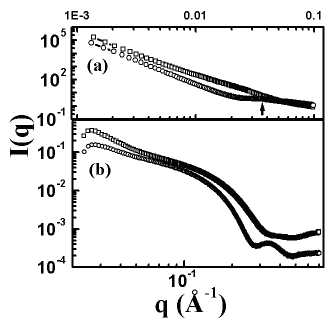}
\caption{ (a): I(q) vs $q$ for  sample A ($\square$)  and B ($\bigcirc$).The location of the structure factor peak is indicated by a vertical arrow.(b) SAXS data for the same samples at higher wave vectors.}
\label{fig:fig1}
\end{figure}
Thus the two samples not only had very different grafting densities but widely different T$_g$ as well. More importantly, the T$_g$ of sample A was higher thatn that of neat PST while that of sample B was lower. We had found similar behaviour in a different system earlier[3],where we had correlated the observed T$_g$ variation to an effective interface morphology parameter. We also found that the microscopic dynamics in two such samples (gold-polymethylmethacrylate) which had widely varying interface morphology and T$_g$ were quite different [29]. However the present system resembled polymeric micelles more closely with the additional degree of control of the core-corona morphology (keeping core size almost fixed) which is suitable for investigating the role of interface morphology of corona on the microscopic dynamics at fixed volume fraction.
We first present results on the detailed structural characterusation of these two sample, especially their respective interface morphology. Fig. 1(a) shows SAXS measurements on samples A and B obtained by
temporal averaging of respective XPCS data. Sample B shows a weak hump at
$q$$\sim$ 0.035 $\AA^{-1}$ and $q$$\sim$ 0.06 $\AA^{-1}$. These peaks are absent in sample A indicating that  it is more polydisperse than sample B. Careful analysis of the SAXS data for sample B in the low $q$ range reveals the presence of fairly monodisperse micellar nanoparticles with a core radius of $\sim$ 16 $\AA$ and a shell of thickness $\sim$ 150  $\AA$.  This clearly indicates that the PST in the shell is highly stretched (radius of gyration, $R_g$, of neat PST is $\sim$ 6 nm) and confirms the high grafting density numebrs we have mentioned earlier and obtained from other estimates. In  Fig. 1(b) we also present the SAXS data collected on the same samples at higher $q$ values in beam line 12-ID of the APS. The shape of the curves are clearly indicative of the sphere form factor and detailed analysis using a Schultz polydisperse sphere form factor leads to gold core radius of 1.6 $\pm$ 0.2 nm and 1.2 $\pm$ 0.3 nm for sample B and A respectively. It is clear that the micellar PST capped gold nanoparticles in
sample B are more monodisperse than those of sample A and are larger
as well. Since the grafting density is higher for nanoparticles in B
than in A the chains of two neighboring particles can not
interpenetrate and hence we have a system which is close to a hard
sphere. Sample A on the other hand has smaller grafting density and
hence a greater probability for chain on neighboring particles to
interpenetrate leading to a sticky hard sphere system. This also
makes the system more disordered. However this difference in
morphology also leads to significant difference in the
dynamics of these systems, as discussed below. 
\begin{figure}
\includegraphics[scale=2]{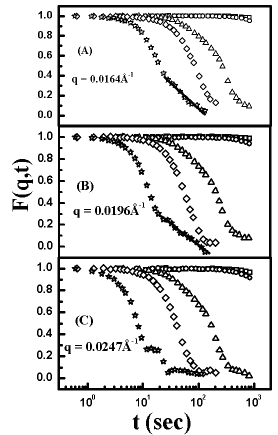}
\caption{  ISF for sample A at respective wave vector, $q$, for different temperatures ($\square$  T = 368K
,$\bigcirc$ T = 378K , $\bigtriangleup$ T = 388K , $\diamond$  T =398K  and $\star$ T = 408K ). The solid lines in panel (a) and (b) indicates the region of logarithmic relaxation.}
\label{fig:fig2}
\end{figure}
It might be mentioned here that the significant difference in glass transition
temperatures between sample A and B can also be explained by the
model for conformation of grafted PST chains on the nano particle
surface which can be modeled as polymer brushes on curved surfaces.
It has been shown earlier [21,27] how by adjusting the grafting density
of capping chains and molecular weight of matrix chains it is
possible to control the graft chains-homopolymer interaction and
hence the thermo-mechanical properties of the polymer nanocomposite
(PNC). It was observed that a wetting interface for lower grafting
density leads to enhancement of T$_g$ of PNC with respect to neat
matrix polymer while a non-wetting interface leads to a reduction in
T$_g$. This is precisely what we also observe for our samples. Fig(2) shows ISF for sample A at 5 different measured temperatures and for various values of wave vector $q$. At temperatures below the measured T$g$ the ISF hardly decays in the measured time window and is independent of $q$. This is a strong signature of an attractive glassy system [12,14-18]. The ISF starts to decay for measured temperatures above the T$_g$ (380 K). 
At intermediate temperatures (T = 388K and 398K), the relaxations look featureless especially at low $q$ , typical of a liquid. However increasing the temperature further to 408K one finds interesting  behavior. It seems clear that a linear region in the relaxation appears,  as indicated by solid lines,at lower $q$. The extent of this region increases further with increase in $q$ till a finite plateau appears at  higher $q$. This two step relaxation is typical of repulsive glasses,the plateau region being indicative of caging. The extent of this plateau is not as long as is typically observed in other system but is indicative of a short-lived dynamic caging that appears in this system at high temperature. The temperature and wave vector dependent behavior is very similar to re - entrant glass transition seen in related systems, where the two glassy phases are separated by a liquid with logarithmic relaxation. If one compares the ISF as obtained for sample B for the same set of measured temperatures, as shown in fig(3), one finds striking differences in behavior. Firstly even at 368 K,the ISF shows decay unlike the behavior for sample A. Secondly, there is no clear evidence of a logarithmic relaxation although a glassy plateau is clearly seen at 408K which is similar to that seen in sample A. In the following we discuss the possible reasons for the striking differences in morphology and dynamics in the two sample PST capped gold nanoparticles with almost identical core sizes. 
\begin{figure}
\includegraphics[scale=2]{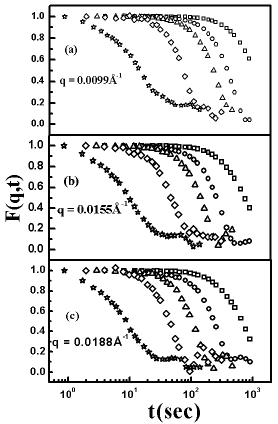}
\caption{ ISF for sample B at respective $q$ for different temperatures ($\square$  T = 368K , $\bigcirc$  T = 378K ,  $\bigtriangleup$ T = 388K , $\diamond$  T = 398K  and  $\star$  T = 408K ). The onset of a plateau corresponding to capping is clearly visible.}
\label{fig:fig3}
\end{figure} 
As mentioned earlier ,the difference in dynamics is due essentially, to the difference in morphology of the capping PST. The longer capping density for sample A allows for inter - digitation between chains of neighboring particles and hence at low temperatures forms a sticky sphere system and hence the dynamics is almost arrested on the time scale of measurement. The larger grafting density allows for variation of the conformation of PST chains on each particle and hence the effective corona thickness has a distribution leading to variation in interparticle spacing between the micellar nanoparticle and hence no clear signature of a peak in SAXS data. For sample B, the higher grafting density leads to strong repulsion or dewetting of chains from neighboring nanoparticles leading to development spatial correlations between the micelles and hence the appearence of the peak at is 0.03 $\AA$$^{-1}$ in SAXS. Thus one can conclude that sample A behaves as a sticky hard sphere system while sample B is closer to a hard sphere system. At higher temperature (408K) the similarity of relaxation between sample A and B does suggest that the capping chains do undergo some reorganization to prevent interparticle chain overlap leading to appearance of effective intrinsic repulsive interaction in the former. We also show how one can trigger an attractive to repulsive glass transition in a micellar nanoparticle - polymer composite system by changing temperature. At the same time by controlling the surface density of grafted polymer chain on nanoparticle surface it is possible to drive the system away from an attractive glassy state to a state resembling a gel. More over we also show for the first time the cross over from the intermediate logarithmic relaxation behavior to a caging dominated plateau in relaxation by changing the wave vector at a fixed temperature and volume fraction. We are making further measurements to map out the phase behavior of such systems but a combined theoretical and experimental effort is required to explore the full phase space of the microscopic dynamics of this very interesting sytem.\\
The authors acknowledge A. Sandy (APS) for assistance in experiments and S.K Sinha (UCSD), L.B.Lurio (NIU) for discussions.This work benefitted by the use of facilities at APS, which is supported by U.S. DOE (BES) under Contract No. W-31-109-Eng-38 to the University of Chicago.Part of the work has been supported by DST, India and UCSD.


\end{document}